     [1999/12/01 v1.4c Il Nuovo Cimento]
\documentclass{cimento}

\usepackage{graphicx}  
\title{GRB follow-up observations in the East-Asian region}
\author{Y.~Urata\from{ins:a}\from{ins:b}\ETC,
K.Y.~Huang\from{ins:c},
W.H.~Ip\from{ins:c},
Y.~Qiu\from{ins:d},
J.Y.~Hu\from{ins:d},
Xn.~Zhou\from{ins:d}
T.~Tamagawa\from{ins:a}
K.~Onda\from{ins:a}\from{ins:e}
\atque
K.~Makishima\from{ins:a}\from{ins:f}
}
\instlist{
\inst{ins:a} RIKEN (Institute of Physical and Chemical Research), 2-1 Hirosawa, Wako, Saitama 351-0198, Japan
\inst{ins:b} Department of Physics, Tokyo Institute of Technology, 2-12-1 Oookayama, Meguro-ku, Tokyo 152-8551, Japan
\inst{ins:c} Institute of Astronomy, National Central University, Chung-Li 32054, Taiwan, Republic of China
\inst{ins:d} National Astronomical Observatories, Chinese Academy of Sciences, Beijing 100012, China
\inst{ins:e} Department of Physics, Tokyo University of Science, 1-3 Kagurazaka, Shinjyuku-ku, Tokyo, Japan
\inst{ins:f} Department of Physics, The University of Tokyo, 7-3-1 Hongo, Bunkyo-ku,Tokyo 113-0033, Japan
}
\PACSes{\PACSit{95.55.Cs}{Ground-based ultraviolet, optical and infrared telescopes} 
\PACSit{98.70.Rz}{ gamma-ray source; gamma-ray bursts}}

\begin{document}

\maketitle

\begin{abstract}

  In 2004, we established a Japan-Taiwan-China collaboration for GRB
study in the East-Asian region. This serves as a valuable
addition to the world-wide optical and infrared follow-up network,
because the East-Asia region would otherwise be blank. We have been
carrying out imaging and spectroscopic follow-up observations at Lulin
(Taiwan), Kiso (Japan), WIDGET (Japan) and Xinglong (China). From
Xinglong and Kiso, we can locate candidates and obtain early time
spectra for afterglows. While WIDGET provides early time observations
before the burst, the high-time resolution for multi-band light curves
can be obtained at Lulin. With the data from these sites, we can
obtain detailed information about the light curve and redshift of
GRBs, which are important to understand the mechanism of the
afterglows.

  Up to March 2005, ten follow-up observations have been provided by
this East-Asia cooperation. Two optical afterglows were detected, GRB
040924 and GRB 041006. The results of the two detected afterglows are
reported in this article.

\end{abstract}

\section{Introduction}

  In East-Asia, since 2004, GRB optical follow-up observations have
been carried out by the cooperation at Japan, Taiwan and China. Due to
the nature of the GRB detection, the Target of Opportunity (TOO)
programs have been provided by the Lulin (Taiwan), Kiso (Japan) and
Xinglong (China) observatories. Since there was a blank in the east
Asia region, the follow-up observations are expected to provide
valuable observations for GRB fields. The position and different
advantages of each site, reduce the risk of bad weather, allowing us
to cover an observational range to up Dec $\sim 40$ degrees, and
provide a complete light curve in multi-wavelengths. Moreover, the
2.16m telescope at Xinglong Observatory is able to perform spectral
observations quickly, as soon as the optical afterglows have been
located. By monitoring the observational ranges of {\it HETE-2} and
{\it Swift}, {\it WIDGET} is capable of detecting the early optical
emission of GRBs.

\section{Site information}
\subsection{Kiso Observatory}
  The TOO system has been prepared for prompt GRB follow-up
observations at Kiso observatory since 2001 \cite{urata}.  The Kiso
observatory is located in Nagano-Prefecture, Japan, and has a 105 cm
Schmidt telescope and two other instruments. One is an optical
2k$\times$2k CCD camera, the other is a near infrared camera named
KONIC (Kiso Observatory Near-Infrared Camera). The 2k$\times$2k
camera's FOV. is 50 $\times$50 arcmin and its limited magnitudes are
22.0 mag, 22.5 mag, 21.0 mag, 21.0 mag. (for each B,V,R,I band, 10
$\sigma$, 900 sec exposure). The KONIC's FOV. is 20 arcmin. There are
also two objective prisms, which allow low-dispersion slit less
spectroscopy with 2k $\times$ 2k CCD with the FOV. 50 $\times$50
arcmin.
For GRB021004, we could obtain the early afterglow spectrum using the
low dispersion slitless spectrograph with an objective prism, well
before an optical transient position was reported \cite{urata03}.

\subsection{WIDGET}

WIDGET is a robotic telescope that monitor the {\it HETE-2} and large
part of the {\it Swift} field of view, to detect optical flashes or
possible optical precursors \cite{tamagawa}. Since WIDGET has a
$62^{\circ}\times62^{\circ}$ F.O.V., we would uniformly obtain optical
light curves before GRBs.
          
\subsection{Lulin Observatory}

The Lulin GRB program based on the Kiso GRB optical observation system
\cite{urata}, started in July 2003 using the Lulin One-meter Telescope
(LOT) in Taiwan. The advantage of the LOT that it is able to perform
small cadence $BVRI-$band time series photometry. The position of
Lulin allows for further the observations of the southern
sky. Moreover, with its better seeing and weather conditions, Lulin is
capable of performing more GRB observations than Kiso either or
Xinglong
\cite{Huang}.

\subsection{Xinglong station of Beijing Observatory}

Bejing observatory's Xinglong station is located in Xinglong County,
Hebei Province, China. The brightness of the sky background on a
moon-less night is $B$ = 22.2, $V$ = 21.0, $R$ = 20.3 , $I$ = 18.8
$\it{mag}$ $\it{arcsec^{-2}}$, respectively. Two telescopes
participate in the East-Asia GRBs cooperation. One is a 2.16$-$m
telescope, which is able to perform optical and near-infrared
low-dispersion spectral observations for GRB afterglows by using the
Optic-Metrics Research spectrograph with Tektronix 1024$\times$1024
CCD. Typically there are about 210 spectroscopic night per year .
The other is the 80$-$cm Cassegrain telescope, which is capable of
automatically response to gamma-ray burst alerts to perform
$BVRI-$band follow-up observations. The F.O.V. is about $ 10' \times
10'$.
      


\section{Current Results}

Up to March 2005, ten follow-up observations have been provided by the
Ease-Asia GRB observations in response to the alerts from {\it
HETE-2}, {\it INTEGRAL} and {\it Swift}. We could provide the upper
limits of magnitude for eight events. Two optical afterglows were
detected, GRB 040924 and GRB 041006. The results for the two detected
afterglows will be described in the following.

\subsection{GRB 040924}

The optical afterglows of short GRBs were elusive until the detection
\cite{Fenimore} of GRB 040924 by {\it HETE-2} on 2004 Sep. 24, at 11:52:11 UT. This event lasted
about 1.2~s and was X-ray rich according to the {\it HETE-2} flux in
the 7--30~keV and 30--400~keV bands. The Konus-Wind satellite also
detected this event with 1.5~s duration in the 20--300~keV band
\cite{Golenetskii}. Since the high-energy spectrum of GRB 040924 is
soft \cite{Fenimore} , the object might actually be near the
short-duration end of the long GRBs. The $R_c$-band and $V$-band
measurements were made at the Luln observatory between 3 and 9 hours
after the burst.  Our optical afterglow observations show that the
temporal evolution, power-law index, and $V-R$ color of GRB 040924 are
also consistent with those of well-observed long GRBs. The signature
of a break in the light curve, as suggested by our present data, can
be explained by the afterglow model, by invoking a early cooling break
\cite{Huang2}.

\subsection{GRB 041006}

Since it was night time in Japan when the burst occurred. The
follow-up started at Kiso, 0.56 hours after the burst. Since the
report by Costa et al (2004)\cite{Costa} had not yet reached us, we
attended to cover the entire {\it HETE-2} error region. We made 300
sec $B$, $V$, and $Rc$ band imaging observations with exposure. In all
bands, the afterglow was detected clearly. An example of the $Rc$ band
image is shown in figure 1\cite{Urata05}.  As shown in figure 1, the
position of the optical afterglow is close to the standard star field
of SA92 reported by Landolt (1992)\cite{Landolt}. To image optical
afterglow and several standard stars from SA92 simultaneously, we
pointed the telescope 5 arcmin to the east.

Since the weather conditions at Lulin were cloudy at the time of the
burst, we had to wait for the weather to inprove before
retrieve. Meantime, we noticed, from our preliminary Kiso results that
the afterglow evolution in the $B$ and $Rc$ bands was significantly
different. Accordingly, we decided to focus on monitoring in these two
particular bands.  We were finally able to start the afterglow
observation at the Lulin observatory at 7.65 hours after the burst. As
planned in advance, we took 300 sec $B$ and $Rc$ band image exposure.
From 2.25 hours, we also performed $Rc$ band imaging observations at
the Beijing observatory, China. Based on the preliminary $B$ and $Rc$
band photometry results form Kiso, the afterglow was estimated te be
slightly too faint to conduct multi color observations using the
Beijing 60 cm telescope. Therefor, we focused on $Rc$ band monitoring.
Thus, thanks to our east Asian network, we were able to obtain
multi-band information in an early phase with least interruptions, in
spite of rather unstable weather conditions at the individual sites.

The multi-band light curve of the GRB 041006 afterglow is shown in
figure 1 . There, we plot our $B$, $V$ and $Rc$ band
points together with several unfiltered and $Rc$ band points reported
to GCN \cite{Yost, Ayani,Fugazza, Monfardini}. The $B$-band light
curve shows a clear plateau around 0.03 days after the burst, although
its behavior in earlier epochs remains unknown.  The $Rc$ band light
curve shows a hint of plateau, or a possible slope change, around 0.1
days after the burst.  Thus, the $B$, $V$ and $Rc$ band photometric
points obtained by our east Asian cooperation plays an important role
in characterizing the temporal evolution of the afterglow.

As is obvious in figure 1, the $Rc$ band light curve is poorly
described by a single power-law ($\alpha=0.82$, $\chi^{2}/\nu=8.1$
with $\nu$=7). However, the data subsets for $t<0.04$ days and
$t>0.15$ days can be successfully described by separate single
power-law models.  We obtained $\alpha=-0.915\pm0.006$ with
$\chi^{2}/\nu=0.0011$ for $\nu=1$, before 0.04 days, and
$\alpha=-1.102\pm0.022$ with $\chi^{2}/\nu=0.65$ for $\nu=4$, after
0.15 days. Thus, the overall behavior of these multi band light curves
may be understood as the sum of two separate components, one showing a
monotonic decay while the other having a rising and falling phases.

\begin{figure}
\begin{center}
\includegraphics[width=0.78\textwidth]{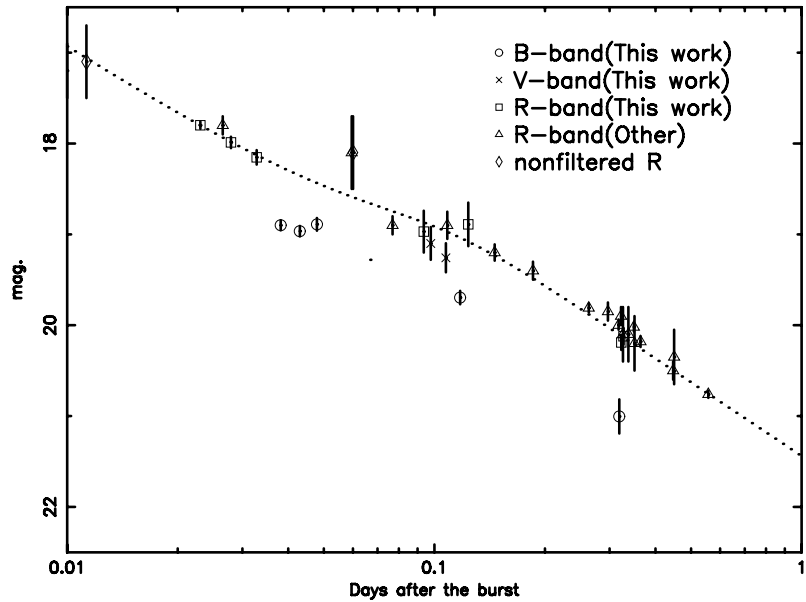}    
\caption{$B$, $V$ and $Rc$ band light curves based on the result at Kiso, 
Lulin and Beijing together with several points reported to GCN circulars.}
\end{center}
\end{figure}

\acknowledgments
We thank the staff and observers at the Kiso, Lulin and XingLong
observatories for the various arrangements. Y.U. acknowledges support
from the Japan Society for the Promotion of Science (JSPS) through a
JSPS Research Fellowship for Young Scientists.

\end{document}